\documentclass{article}

\usepackage{arxiv}

\usepackage[utf8]{inputenc} 
\usepackage[T1]{fontenc}    
\usepackage{booktabs}       
\usepackage{amsfonts,amsmath,amssymb}       
\usepackage{nicefrac}       
\usepackage{microtype}      
\usepackage{lipsum}		
\usepackage{graphicx}
\usepackage{natbib}
\usepackage{doi}

\usepackage{url,hyperref,lineno,microtype,subcaption}
\usepackage[onehalfspacing]{setspace}
\usepackage{algorithmic}
\usepackage{booktabs}
\usepackage{array}
\usepackage{soul,color}

\newcommand{\Un}{U_\mathrm{neg}}
\newcommand{\Up}{U_\mathrm{pos}}
\newcommand{\lampe}{\mathrm{LAM}_\mathrm{PE}}
\newcommand{\lamne}{\mathrm{LAM}_\mathrm{NE}}
\newcommand{\lli}{\mathrm{LLI}}
\newcommand{\xmin}{x_{\mathrm{min}}}
\newcommand{\xmax}{x_{\mathrm{max}}}
\newcommand{\ymin}{y_{\mathrm{min}}}
\newcommand{\ymax}{y_{\mathrm{max}}}
\newcommand{\qli}{Q_\mathrm{Li}}

\newcommand{\nmcone}{\mathrm{Ni}_\mathrm{0.33}\mathrm{Mn}_\mathrm{0.33}\mathrm{Co}_\mathrm{0.33}}

\newcommand{\nmcsix}{\mathrm{Ni}_\mathrm{0.6}\mathrm{Mn}_\mathrm{0.2}\mathrm{Co}_\mathrm{0.2}}

\newcommand{\qsei}{\tilde{Q}_{\mathrm{SEI}}}

\newcommand{\nprp}{\mathrm{NPR}_\mathrm{practical}}

\definecolor{lightblue}{rgb}{0.7,1,1}
\definecolor{lightgray}{rgb}{0.9,0.9,0.9}
\definecolor{lightgreen}{rgb}{0.7,1,0.7}

\newif\ifrespondingtoreviewers
\respondingtoreviewersfalse

\ifrespondingtoreviewers
    \newcommand{\reva}[1]{\sethlcolor{lightblue}\hl{#1}}

    \newcommand{\rev}[1]{\sethlcolor{lightgray}\hl{#1}}
\else
    \newcommand{\rev}[1]{#1}
    \newcommand{\reva}[1]{#1}

\fi

\title{Differential voltage analysis for battery manufacturing process control}

\author{
    \hspace{1mm}Andrew Weng \\
	Department of Mechanical Engineering\\
	University of Michigan\\
	Ann Arbor, MI 48109 \\
	\texttt{asweng@umich.edu} \\
	\And
	\hspace{1mm}Jason B.~Siegel \\
	Department of Mechanical Engineering\\
	University of Michigan\\
	Ann Arbor, MI 48109\\
	\texttt{siegeljb@umich.edu} \\
	\And
    Anna Stefanopoulou \\
    Department ofMechanical Engineering \\
    University of Michigan \\
    Ann Arbor, MI 48109 \\
    \texttt{annastef@umich.edu}
}

\renewcommand{\shorttitle}{Differential Voltage Analysis for Battery Manufacturing Process Control}

\hypersetup{
pdftitle={Differential voltage analysis for battery manufacturing diagnostics},
pdfsubject={q-bio.NC, q-bio.QM},
pdfauthor={Andrew Weng, Jason B.~Siegel, Anna Stefanopoulouum},
pdfkeywords={First keyword, Second keyword, More},
}

\begin{document}
\maketitle

\begin{abstract}

Voltage-based battery metrics are ubiquitous and essential in battery manufacturing diagnostics. They enable electrochemical ``fingerprinting'' of batteries at the end of the manufacturing line and are naturally scalable, since voltage data is already collected as part of the formation process which is the last step in battery manufacturing. Yet, despite their prevalence, interpretations of voltage-based metrics are often ambiguous and require expert judgment. In this work, we present a method for collecting and analyzing full cell near-equilibrium voltage curves for end-of-line manufacturing process control. The method builds on existing literature on differential voltage analysis (DVA or dV/dQ) by expanding the method formalism through the lens of reproducibility, interpretability, and automation. Our model revisions introduce several new derived metrics relevant to manufacturing process control, including lithium consumed during formation and the practical negative-to-positive ratio, which complement standard metrics such as positive and negative electrode capacities. To facilitate method reproducibility, we reformulate the model to account for the ``inaccessible lithium problem'' which quantifies the numerical differences between modeled versus true values for electrode capacities and stoichiometries. We finally outline key data collection considerations, including C-rate and charging direction for both full cell and half cell datasets, which may impact method reproducibility. This work highlights the opportunities for leveraging voltage-based electrochemical metrics for online battery manufacturing process control.

\end{abstract}


\newpage

\begin{figure*}[h!]
\begin{center}
\includegraphics[width=17.5cm]{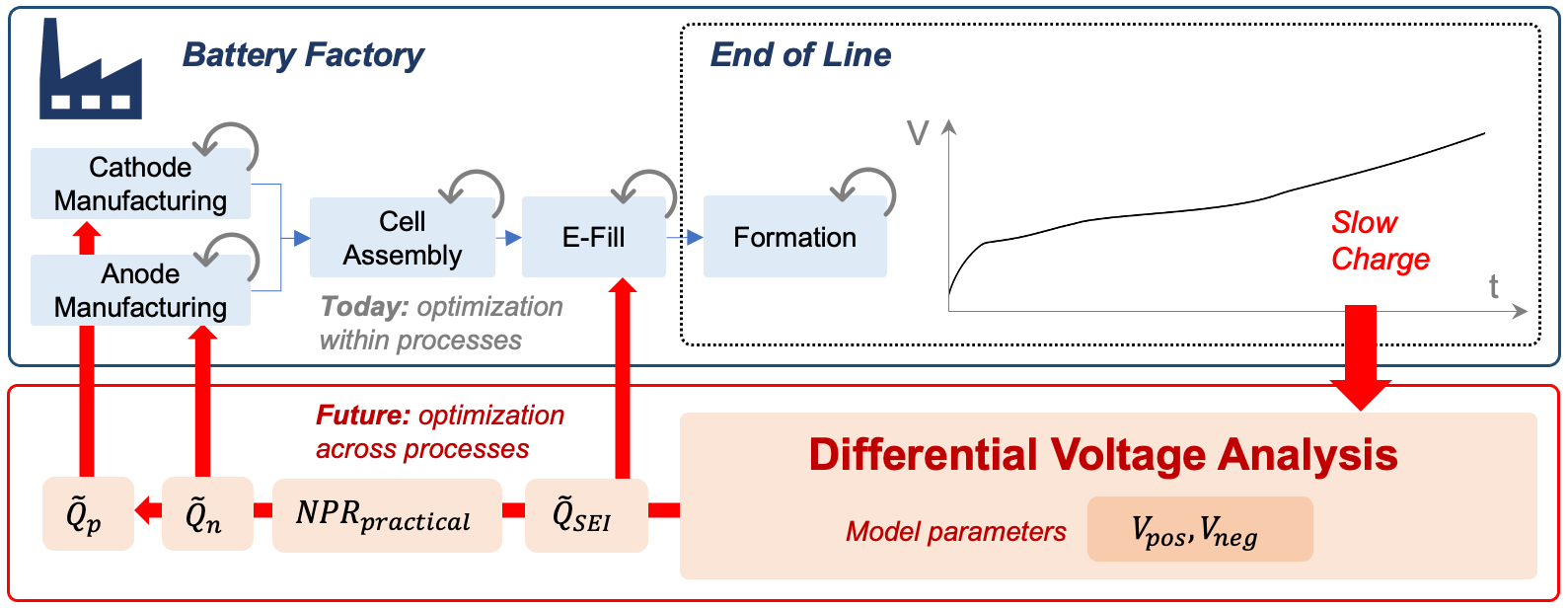}
\end{center}
\label{fig:graphical_abstract}
\centering \textbf{Graphical Abstract}
\end{figure*}

\section{Introduction}

\rev{Decreasing} the cost and environmental \rev{footprint} of battery ``gigafactories'' worldwide relies on continuously improving the manufacturing process through data-driven process control. To place the scale of manufacturing data in context, consider a battery factory that supplies enough cells to produce one electric vehicle (EV) per minute \citep{Kane2022-yg}. Since the number of cells needed for each EV is on the order of hundreds for pouch cells and thousands for cylindrical cells, the cell production rate would need to exceed $\sim$100 cells per minute for pouch cells and $\sim$1,000 cells per minute for cylindrical cells. Every cell produced will generate voltage and current time-series data as part of the formation process, which is the last step of battery manufacturing when cells are charged for the very first time to create the solid electrolyte interphase (SEI) \citep{Peled2017-df, An2017, Wood2019-uu}.

Voltage and current data from the battery formation process can be continuously and automatically collected, stored, and analyzed to develop smart manufacturing process specifications \rev{or} tolerances, ensuring that all cells leaving the factory have a guarantee on performance, lifetime, and safety \citep{Liu2021-ye}. Using voltage-based measurements is appealing from a manufacturing standpoint since they can be measured using existing formation cyclers. \rev{This data may be directly collected as part of the formation protocol or immediately after the formation protocol completes but before the cells are taken off of the formation cyclers} \citep{Weng2021}. Thus, collecting voltage-based data bears no additional \rev{capital} costs to a factory. Since the number of formation cyclers grows with the \rev{production volume}, the voltage-based metrics will also naturally scale as factory throughput increases. 

When \rev{full cell} voltage data is carefully collected and analyzed, they can be used to derive physically-interpretable, electrochemical \rev{metrics, or features,} that provide insights into the thermodynamic and kinetic properties of the cell. A \rev{s}low-rate \rev{(C/20) full cell} voltage curve can, for example, be analyzed using incremental capacity analysis (ICA) \citep{Dubarry2022-ax} or differential voltage (dV/dQ) analysis \citep{Dahn2012-nz, Lee2020a} to understand thermodynamic cell properties such as active material losses and lithium inventory losses. \rev{The data can be collected immediately after the formation cycles complete.} Unlike cell dissections and electrode harvesting, voltage-based analysis is non-destructive. \rev{The same cell that is analyzed in manufacturing can thus be tracked throughout its remaining life, either through accelerated cycle life testing in a lab or through fleet telemetry of real-world usage inside of an EV}. The ability to establish an electrochemical ``fingerprint'' on the pristine cells immediately after manufacturing can help improve lifetime prediction models \citep{Weng2021} which, in turn, be can be used to improve the manufacturing process.

However, caution is needed when setting manufacturing specifications based on voltage-based data. Without careful data collection and interpretation, manufacturing \rev{tolerances} may be set \rev{too tightly, increasing reject rates and lowering production throughput, or too loosely, which may increase production throughput in the short term but lead to lifetime and safety issues after years of usage in the field.} Ultimately, manufacturing specifications must \rev{also} be set based on an understanding of long-term consequences to performance, lifetime, or safety. The voltage-based electrochemical features can facilitate understanding \rev{trade-offs between production throughput and long-term reliability}, but only if they are carefully designed and interpreted.

We note that voltage-based electrochemical features do not replace the need for more advanced end-of-line metrology methods in the factory, including X-ray \citep{Kong2022-gl, Qian2021-oi} and ultrasonic imaging \citep{Bommier2020-so, Deng2020-kd}, which may be necessary for catching non-electrochemical related cell defects. Rather, the voltage-based features complement root-cause analyses by providing basic electrochemical \rev{metrics} about the system such as the as-manufactured electrode capacities, cyclable lithium inventory, and negative-to-positive ratio (NPR).

Currently, factories are not taking full advantage of voltage-based measurements \rev{at the end of line which may be relevant to long term cell lifetime and durability, partly owing to difficulties in data interpretation.} Resolving the complex electrochemical details using full-cell voltage data alone is inherently challenging, since these measurement\rev{s} reflect a combination of thermodynamic and kinetic factors originating from nearly every cell component \rev{and their interactions} with each other. These inherent difficulties are compounded by a \rev{general} lack of clear literature guidance on best practices for collecting and analyzing voltage data. As a result, commonly reported electrochemical metrics derived from voltage data, such as modeled electrode capacities from dV/dQ analysis often differ in both data collection and analysis methods, limiting their interpretability and reproducibility. \rev{To overcome these challenges, recent work, such as those by} \cite{Dubarry2022-ax} \rev{are essential for enabling reproducible extraction of electrochemical features derived from voltage data.}

The goal of this work is thus to guide experimental and analysis considerations needed to enable reproducible voltage-based battery manufacturing diagnostics. To accomplish this, we revisit a popular analysis technique through the lens of battery manufacturing: the \rev{differential voltage analysis, or dV/dQ, model.} We will detail the experimental and analysis considerations to facilitate data reproducibility and interpretability.

\section{A Manufacturing Case Study}
\label{sec:application1}

Here, we showcase an example of how \rev{the differential voltage analysis method can be used for continuous monitoring of cell electrochemical features at the end of the manufacturing process. We highlight the various electrochemical metrics, or features, that can be extracted using the method and their significance for manufacturing process control.} 

The study consists of two cell batches made on the same manufacturing line and made using the same active materials ($\nmcone$ positive electrode and graphite negative electrode). However, the two cell batches were made two years apart and differed in cell design parameters, including electrode loading targets and number of layers. Cells from the first batch, from \cite{Mohtat2021b}, had nominal capacities of 5.0 Ah and were built in 2018 \rev{($n=21$)}. Cells from the second batch, from \cite{Weng2021}, had nominal capacities of 2.37 Ah and were built in 2020 \rev{($n=40$). For these cells, we restricted the cell population to those having full cell data collected at room temperature (RT) ($n=20$) to match the measurement conditions from} \cite{Mohtat2021b}. A table comparing relevant design parameters is given in Table S1. We will refer to these two datasets as Mohtat2021 and Weng2021, respectively.

\begin{figure}[h!]
\begin{center}
\includegraphics[width=17.5cm]{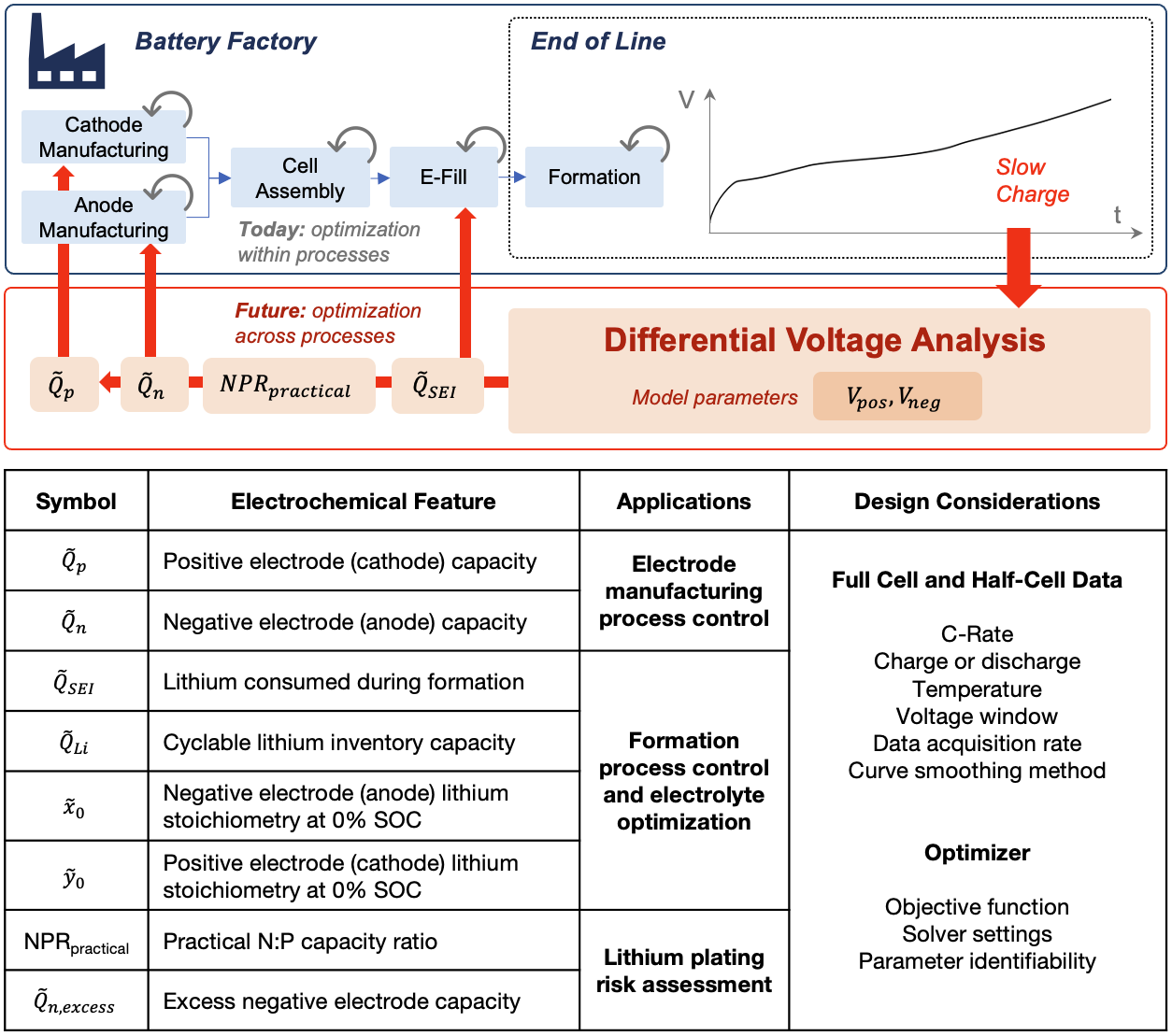}
\end{center}
\caption{\textbf{Summary of how differential voltage analysis can enable coordinated battery manufacturing process control via end-of-line testing}. The provided table describes the voltage-based electrochemical features derived from the method, including applications in manufacturing and test design considerations.}
\label{fig:features}
\end{figure}

\rev{Figure }\ref{fig:features} \rev{shows the general analysis workflow and lists the electrochemical features extracted by the model. In this example, full cell voltage data ($V_\mathrm{full}$)} was collected on every manufactured cell after the formation cycling completed (Section \ref{sec:fullcell}). Each dataset was analyzed using the differential voltage analysis model (Section \ref{sec:voltage-fitting-model}). The model consisted of two pre-computed functions, $\Up$ and $\Un$, representing the positive and negative electrode near-equilibrium reference potential curves, respectively (Section \ref{sec:halfcell}). The same functions were used to analyze data from both Mohtat2021 and Weng2021, since the batches shared the same positive and negative electrode chemistries.

\rev{The differential voltage model was then used to extract electrochemical features which include positive and negative electrode capacities ($\tilde{Q}_p$, $\tilde{Q}_n$), the capacity of lithium available for cycling ($\tilde{Q}_\mathrm{Li}$), the capacity of lithium lost to the SEI ($\tilde{Q}_\mathrm{SEI}$), electrode lithium stoichiometries when the full cell is discharged ($\tilde{x}_0$, $\tilde{y}_0$), and cell design information including the practical negative-to-positive capacity ratio ($\mathrm{NPR}_\mathrm{practical}$) and the excess negative electrode capacity ($\tilde{Q}_{n,\mathrm{excess}}$). The `tildes' on each variable name indicate that they are estimates which may be different from the true parameters due to the ``inaccessible lithium problem,'' which will be further explained in Section} \ref{sec:inaccessible}. These extracted features can then be analyzed to determine whether machine parameter adjustments are needed in upstream manufacturing processes such as those in electrode coating or in electrolyte fill (E-fill). 

\begin{figure}[h!]
\begin{center}
\includegraphics[width=14.5cm]{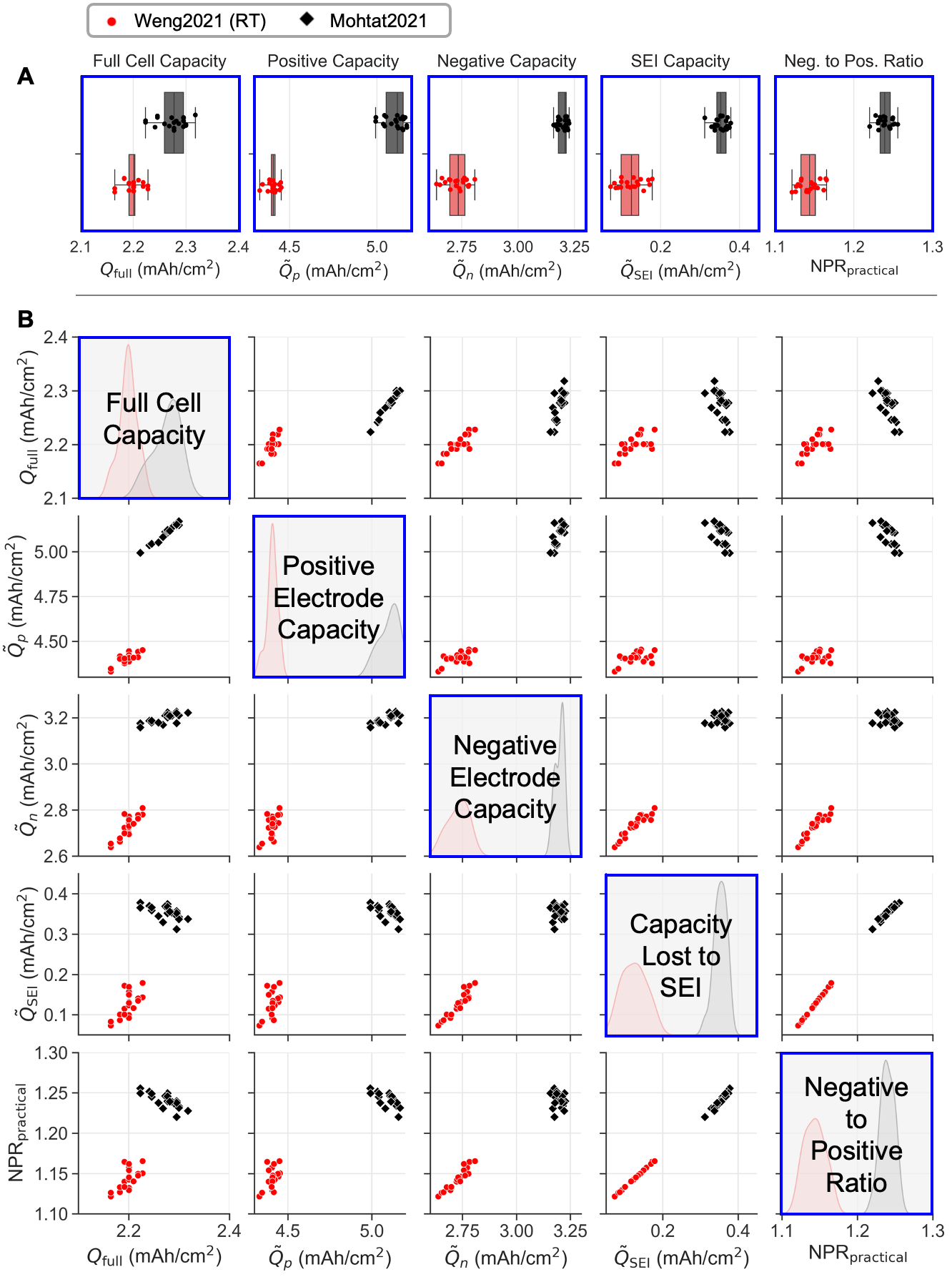}
\end{center}
\caption{\textbf{Leveraging differential voltage analysis to non-destructively identify electrochemical differences between two cell batches at the end of the manufacturing line.} $Q_\mathrm{full}$: full cell areal capacity, $\tilde{Q}_p$: \rev{positive electrode} areal capacity, $\tilde{Q}_n$: \rev{negative electrode} areal capacity, $\tilde{Q}_\mathrm{SEI}$ areal capacity of lithium lost to SEI during formation, $\mathrm{NPR}_\mathrm{practical}$: Practical negative to positive ratio. Data spans two separate datasets from \cite{Weng2021} \rev{($n=20$)} and \cite{Mohtat2021b} \rev{($n=21$)}. Cells from both datasets share the same chemistry ($\nmcone|$graphite), but were made in two separate batches and vary in cell design parameters. (A) Box-and-whisker plots comparing data from the two batches, where each marker shows modeled outputs on each individual cell. (B) correlation plots of the metrics, where diagonal elements show histograms comparing individual metrics across the two different batches. }
\label{fig:correlations-manufacturing}
\end{figure}

Figure \ref{fig:correlations-manufacturing} show an example data visualization of the analysis outputs. An example model fitting result is also provided in Figure S1. Panel A uses box-and-whisker plots to compare compares several example metrics derived from the differential voltage analysis, including \rev{$\tilde{Q}_p, \tilde{Q}_n$, $\tilde{Q}_\mathrm{SEI}$, and $\mathrm{NPR}_\mathrm{practical}$}. The full-cell C/20 discharge capacity, $Q_{\mathrm{full}}$, is also provided for reference. Since the Mohtat2021 and Weng2021 cells used different numbers of electrode sheets, all of the capacity values reported here have been normalized to the respective electrode areas, \rev{enabling a head-to-head comparison.}

\rev{The following sections discusses how to interpret each group of features in the context of manufacturing process control.}

\subsection{Monitoring Electrode Capacities and Loadings}
\label{sec:loadings}

Figure \ref{fig:correlations-manufacturing}A \rev{shows extracted values of positive and negative electrode capacities, $\tilde{Q}_n$ and $\tilde{Q}_p$, for every cell in this study. In a manufacturing context, the availability of data on $\tilde{Q}_n$ and $\tilde{Q}_p$} on every batch of cells in production, without needing cell dissections, can help to identify deviations from electrode loading set-points, e.g. due to electrode processing variations. These electrode-level variations would be difficult to detect by relying on full cell capacity check data alone, since capacity variability at the electrode level may not manifest in full cell capacity checks (Section \ref{sec:variability}).

We also found that the Mohtat2021 cells had, on average, 16\% higher \rev{positive electrode} areal capacity and 17\% higher \rev{negative electrode} areal capacity compared to the Weng2021 cells. \rev{However, according to the cell design parameters (Table S1), Mohtat2021 cells should have only 7\% higher positive electrode areal capacity and 9\% higher negative electrode areal capacity based on differences in the active material loading targets. The gap between the the targeted and modeled capacities suggest that the as-built cells did not match their target loadings, and that electrode coating process parameters may need to be adjusted.}

\subsection{Monitoring Lithium Consumed During Formation}
\label{sec:lithium-sei}

Figure \ref{fig:correlations-manufacturing}A also compares the amount of lithium consumed during formation, $\tilde{Q}_{\mathrm{SEI}}$, for all cells in this study. For the Mohtat2021 cells, the mean value was measured to be 0.35 mAh/cm$^2$, while the Weng2021 cells measured 0.12 mAh/cm$^2$. Mohtat2021 cells thus lost nearly three times more lithium during formation than compared to the Weng2021 cells. \rev{This result can be verified graphically by inspecting the voltage curve alignments from the fitting results which shows that the positive electrode is less lithiated at 0\% SOC, reflecting the fact that extra lithium was irreversibly consumed to form the SEI and cannot return to the positive electrode on discharge (Figure S1).} The increase in lithium consumption rate during the formation of the Mohtat2021 cells suggests a poorer quality of electrolyte or a less passivating \rev{negative electrode} formulation. Note that both cell builds shared the same formation protocol, so differences in lithium consumption cannot be attributed to differences in the formation protocol. \rev{This example highlights a scenario in which the electrolyte filling process parameters may need to be inspected, or the quality of the electrolyte itself may need to be verified.}

\rev{The increased lithium consumption in the Mohtat2021 cells also helps to explain why, although the Mohtat2021 cells had 16\% higher positive electrode areal mass loading than the Weng2021 cells, the full cell capacities of the Mohtat2021 cells were only 3\% higher on average. Thus, although the Mohtat2021 cells were designed with higher areal capacities than the Weng2021 cells, most of the benefits to full cell capacity were lost due the increased lithium consumption rate during formation.}

\subsection{Measuring ``Margin to Lithium Plating'' Using the Practical NP Ratio}
\label{sec:plating-margin}

Since the Mohtat2021 cells had consumed more lithium during formation compared to the Weng2021 cells, the Mohtat2021 \rev{negative electrodes} will also be less lithiated when the full cell is charged to 100\% SOC. The Mohtat2021 cells \rev{will therefore} have a wider margin of \rev{negative electrode} capacity before the \rev{negative electrode} is fully lithiated and lithium plating begins to occur, a result that can be visually seen in Figure S1. Thus, Mohtat2021 cells should, \rev{in theory}, be less prone to lithium plating compared to the Weng2021 cells. 

A metric that quantifies the capacity margin before lithium plating occurs is the negative to positive ratio (NPR). The differential voltage analysis provides such a metric, known as the practical NPR, or $\mathrm{NPR}_\mathrm{practical}$, which we develop in Section \ref{sec:npr}. The $\nprp$ for Mohtat2021 cells was measured to be 1.24, compared to 1.14 for the Weng2021 cells, confirming that the Mohtat2021 cells have a higher \rev{negative electrode} capacity margin to protect against lithium plating. In order for the Weng2021 cells to achieve the same $\nprp$ as the Mohtat2021 cells, either the positive electrode loading target would need to decrease, or the negative electrode loading target would need to increase. 

\subsection{Electrode Manufacturing Variability}
\label{sec:variability}

\rev{Turning to Figure} \ref{fig:correlations-manufacturing}B, we next show how the differential voltage analysis outputs can be used to identify differences in electrode manufacturing variability from batch to batch. \rev{W}e immediately see that Mohtat2021 cells showed higher variability in the full cell capacity. The histograms of \rev{$\tilde{Q}_n$ and $\tilde{Q}_p$} reveal that the origin of the higher full cell capacity variability stems from the \rev{positive electrode} loading, not the \rev{negative electrode} loading. In fact, the \rev{negative electrode} loading for Weng2021 cells had higher variability compared to the Mohtat2021 cells, yet this increased variability did not manifest in the full cell capacity. This result is expected considering the fact that the practical NP ratio of all cells are greater than 1\rev{, and hence, variations in the negative electrode capacities will mainly manifest in the excess negative electrode capacity rather than the full cell capacity.} Thus, to decrease the full cell capacity variability in the Mohtat2021 cells, the manufacturer should focus on tightening the variability in the cathode manufacturing process, not the anode manufacturing process.

\subsection{Parameter Correlations}

Figure \ref{fig:correlations-manufacturing}B \rev{finally presents correlation plots between every individual output variable of the differential voltage analysis model.} One application of studying correlations is to understand which electrode-level parameters correlate to full cell capacity and hence \rev{directly impacts} measurable cell performance. The correlation matrix reveals that, \rev{within each batch of cells, the positive electrode} capacity correlates to full cell capacity. However, the correlation breaks when comparing across different batches of cells since the Mohtat2021 consumed more lithium during formation (Section \ref{sec:lithium-sei}). This result highlights the reality that multiple cell manufacturing factors, including electrode loadings, lithium consumption during formation, and the \rev{negative electrode} excess capacity, all play a role in determining the full cell capacity. Achieving a certain full cell capacity outcome therefore requires coordinated control across multiple manufacturing processes.

In summary, this section highlighted the utility of applying the \rev{differential voltage analysis method to extract electrochemical metrics at the end of the cell manufacturing line. These metrics can then be directly used to adjust process upstream manufacturing process parameters such as those in electrolyte manufacturing or electrolyte filling.} All of the data presented in this section were directly outputted from the \rev{differential voltage method} without needing any electrode-level degradation analysis. Since the method can run on data collected during battery formation, no additional work is necessary to extend the analysis to new cells coming off the manufacturing line, making the method highly scalable.

\clearpage

\section{Differential Voltage Analysis Method}
\label{sec:dva}

\subsection{Model Selection}

Near-equilibrium (i.e. slow-rate) full cell voltage curves have been widely used to understand dominant cell degradation modes \rev{such as loss of active material (LAM) and loss of lithium inventory (LLI)} \citep{Bloom2005, Dubarry2012, Birkl2017a, Olson2023-yi}. The appeal of these techniques is easy to appreciate: full cell data is easier and faster to collect than materials-level characterization data which often requires cell tear-down and electrode harvesting. \rev{Such types of analyses} have been introduced under different names, including differential voltage analysis (DVA or dV/dQ) \citep{Bloom2005, Smith2011a}, incremental capacity analysis (ICA) \citep{Dubarry2012, Weng2013a, Dubarry2022-ax}, open-circuit voltage models \citep{Birkl2017a, GarciaElvira2021, Schmitt2022-mr}, and voltage-fitting analysis \citep{Lee2020a}. \rev{While all of these methods rely on full cell voltage curve data, their data analysis approaches are different. In differential voltage (or dV/dQ) methods, the voltage is differentiated with respect to capacity and plotted against capacity. The resulting data reveals peaks and troughs that can be attributed to either the positive or negative electrode. In incremental capacity (or ICA) methods, the capacity is differentiated with respect to voltage and plotted against voltage. The features observed from ICA are no longer linearly separable, but through careful data interpretation, distinct degradation modes can still be inferred} \citep{Dubarry2022-ax}. \rev{Other variations of differential analysis have also been recently introduced which leverage signals such as expansion} \citep{Mohtat2022-xi} and pressure \citep{Huang2022-zo}, but since these methods require measurements beyond full voltage, they are omitted from present consideration.

Since the exact method for extracting the electrode-level parameters vary from author to author, comparing results across different papers is complicated. Notably, some authors use full cell data alone, assigning features of interest \rev{and} graphically infer\rev{ring} degradation metrics such as lithium inventory and electrode capacity losses \citep{Dubarry2021}, while others leverage half-cell ``reference curves'' \citep{Harlow2019-et} to build a model of the full cell voltage as a function of the electrode-level parameters \citep{Birkl2017a, GarciaElvira2021}. 

In a battery manufacturing context, a quantitative and reproducible voltage-based analysis method is needed. The method must be quantitative in order to resolve minute differences in manufacturing process parameters which could have large, long-term consequences to battery performance, lifetime and safety. The method must also be reproducible in order to scale. 

Out of all of the \rev{available} voltage curve analysis methods, the ``voltage fitting method'', documented by \cite{Lee2020a}, provides the best balance between mathematical rigor and interpretability. We therefore choose this method as the basis for extracting electrochemical features for manufacturing diagnostics. \rev{Note that, in this work, we use the words ``differential voltage analysis'' and ``voltage fitting analysis'' interchangeably since they refer to the same fundamental model construction.}

\subsection{Model Formulation}
\label{sec:voltage-fitting-model}

The \rev{differential voltage} model seeks to decompose the full cell near-equilibrium \reva{(i.e. open circuit)} voltage curve into its constituent half-cell positive and negative potential curves vs Li/Li$^+$. The half-cell curves, also sometimes called ``reference curves'' \citep{Harlow2019-et}, then provide information about thermodynamic states of the cell\rev{, including positive and negative electrode capacities, positive and negative electrode lithium stoichiometries as a function of the full cell state of charge (SOC), and the capacity of cyclable lithium inventory}.

\begin{figure}[h!]
\begin{center}
\includegraphics[width=17.5cm]{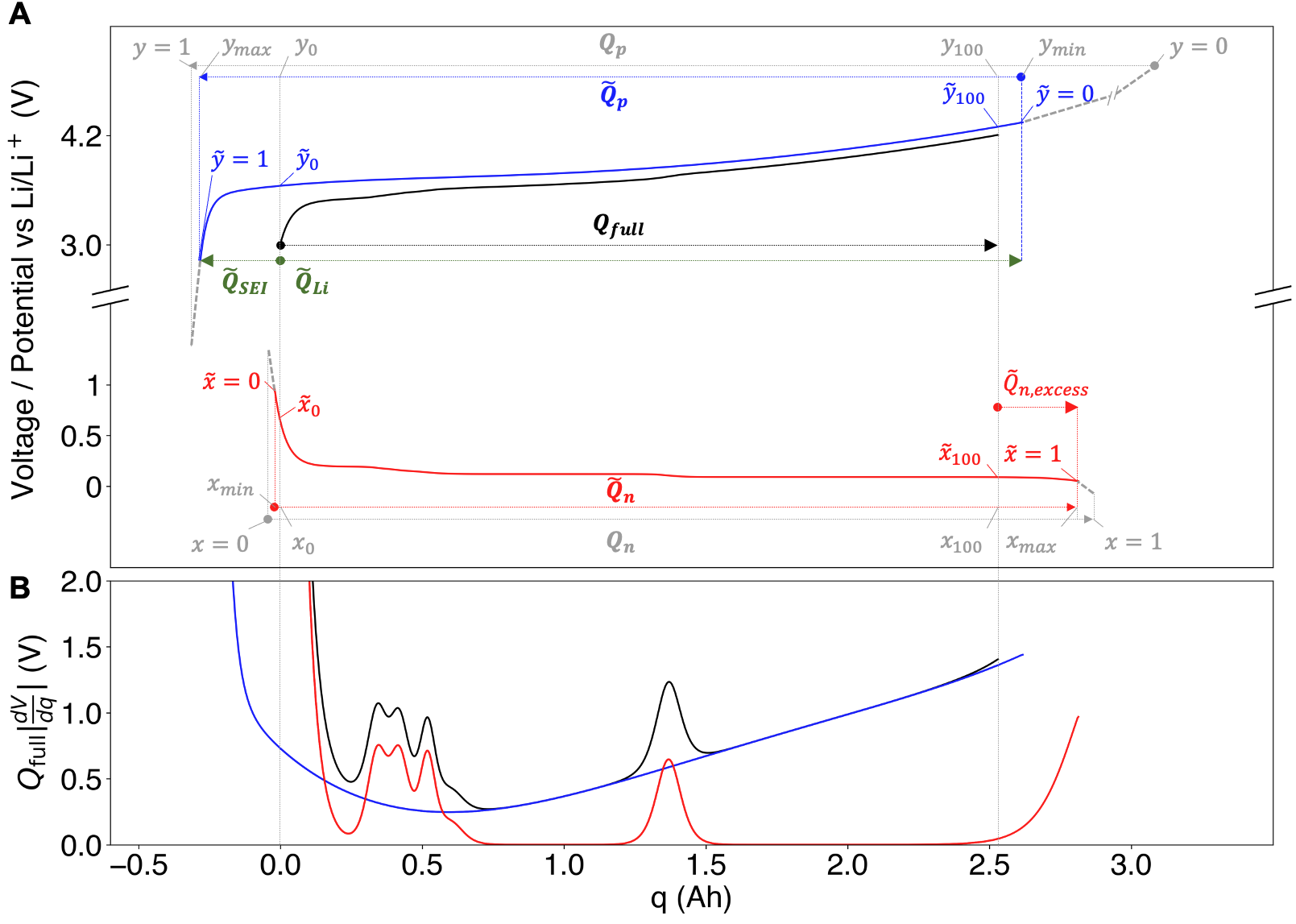}
\end{center}
\caption{\textbf{Concept illustration of the differential voltage analysis method and the inaccessible lithium problem.} (A) \reva{The full cell near-equilibrium (`open circuit') voltage curve $V_\mathrm{full}$ (black) plotted alongside the corresponding positive electrode (blue) and negative electrode (red) near-equilibrium potential curves $\Up$ and $\Un$}. Parameters with tildes indicate estimated model parameters due to the ``inaccessible lithium problem'' (see Section \ref{sec:inaccessible}). Gray lines and variables indicate true model parameters that cannot be ascertained using the native model. (B) The corresponding differential voltage (dV/dQ) curves, illustrating how full cell dV/dQ peaks can be attributed to features from the positive and negative electrodes.}
\label{fig:dvdq}
\end{figure}

We now proceed with a quantitative treatment of the \rev{model, using} Figure \ref{fig:dvdq} \rev{as a visual guide}. The fundamental state equation of the model is an assertion of voltage conservation between the half-cell near-equilibrium potentials and the full cell near-equilibrium potential given by

\begin{equation}
    \label{eq:voltage}
    V_{\mathrm{full}}(q) = \Up(y) - \Un(x),
\end{equation}

where $V_{\mathrm{full}}$ is the full cell \reva{near-equilibrium (i.e. open-circuit)} voltage and $\Up$ and $\Un$ are the positive and negative electrode \reva{near-equilibrium} potential curves versus Li/Li${^+}$, \rev{respectively}. $q$ is a vector of capacities in the full cell domain and is related to the cell SOC, $z \in [0,1]$, by the relation

\begin{equation}
    \label{eq:soc}
    z = \frac{q}{Q_\mathrm{full}}
\end{equation}

where $Q_\mathrm{full}$ is the measured capacity of the full cell. $x\in[0,1]$ and $y\in[0,1]$ are vectors of lithium stoichiometries in the negative and positive electrodes, \rev{respectively}. Note that $x$ and $y$ are sometimes also reported as $\theta_n$ and $\theta_p$ by other authors. The lithium stoichiometries are \rev{governed} by the half-cell reactions

\begin{align}
    x\mathrm{Li^+} + x\mathrm{e^-} + \mathrm{C}_6 \rightarrow \mathrm{Li}_x\mathrm{C}_6 &\qquad\qquad \mathrm{(Negative, charging)} \\ 
    \mathrm{LiM} \rightarrow \mathrm{Li}_y\mathrm{M} + (1-y)\mathrm{Li^+} + (1-y)\mathrm{e^-} &\qquad\qquad \mathrm{(Positive, charging)},
\end{align}

where the first reaction represents the intercalation of lithium into graphite, and the second reaction \rev{describes the} release of lithium from any intercalation positive electrode such as (NMC), nickel cobalt aluminum (NCA), iron phosphate, etc. 

Equation \ref{eq:voltage} assumes that both electrodes are at \reva{near-equilibrium, meaning that the current densities used to acquire the curves are sufficiently small so that over-potential contributions can be neglected.} Note that while some authors attempt to use $V_{\mathrm{full}}$ to extract kinetic information such as resistance increase \citep{Dubarry2012}, we focus our analysis on extracting only thermodynamic quantities to limit the number of modeled output parameters and improve solution uniqueness. To ensure negligible kinetic effects, $V_{\mathrm{full}}(q)$ must be collected at a sufficiently slow rate (see Section \ref{sec:fullcell}). The reference half-cell curves $\Up(y)$ and $\Un(x)$ may be obtained experimentally, e.g. via coin cell \rev{testing} (see Section \ref{sec:halfcell}).

The terms in \rev{E}quation \ref{eq:voltage} can be mapped to a common, full capacity basis $q$. Charge conservation requires that this mapping be affine (linear plus an offset). We can therefore define the following basis transformation equations:

\begin{align}
    \label{eq:x}
    x(q) &= x_0 + \frac{q}{Q_n} \\
    \label{eq:y}
    y(q) &= y_0 - \frac{q}{Q_p},
\end{align}

where $Q_n$ and $Q_p$ are the total negative and positive electrode capacities, respectively, and $(x_0,y_0)$ are the negative and positive electrode stoichiometries in the fully discharged state. These equations are derived under the \rev{boundary condition} that $q = 0$ corresponds to the fully discharged state, with $(x,y) = (x_0,y_0)$. Equations (\ref{eq:x}, \ref{eq:y}) can be inverted to obtain

\begin{align}
    \label{eq:q_x}
    q(x) = Q_n(x - x_0) \\
    \label{eq:q_y}
    q(y) = Q_p(y_0 - y),
\end{align}

which can be used to map lithium stoichiometries to the shared capacity basis. Note that, under this construction, $q$ is allowed to take on negative values. For example, when \rev{$x=0$, $q = -Q_nx_0 < 0$, according to Equation} \ref{eq:q_x}. Negative $q$ values can be interpreted as ``virtual capacities,'' that is, capacity present in the electrodes but is inaccessible from the perspective of the full cell due to the full cell minimum voltage constraint. 

We can also define a complementary set of variables describing the lithium stoichiometries in the fully-charged state, $(x_{100},y_{100})$, which can be written as

\begin{align}
    \label{eq:x100}
    x_{100} &= x_0 + \frac{Q_{\mathrm{full}}}{Q_n} \\
    \label{eq:y100}
    y_{100} &= y_0 - \frac{Q_{\mathrm{full}}}{Q_p}.
\end{align}

$Q_{\mathrm{full}}$ is a known quantity from the full cell data and is not considered to be a modeled parameter. Hence, $(x_{100},y_{100})$ are not \rev{independent} model parameters. 

In summary, knowledge of four parameters

\begin{equation}
    \theta = \{Q_n, Q_p, x_0, y_0\}
\end{equation}

provides a complete description of the model from Equation (\ref{eq:voltage}). These parameters bear physical meaning, since ($Q_n,Q_p$) are related to the total electrode capacities and $(x_0, y_0)$ predict lithium stoichiometry-dependent electrode potentials. These model parameters can then be used to derive additional electrochemical features \rev{relevant to battery manufacturing diagnostics}, as will be described in Section \ref{sec:manufacturing-dvdq}. 

Note that the choice of model parameters are not unique. For example $\{Q_n, Q_p, x_{100}, y_{100}\}$ is an equally valid set of parameters. For the remainder of this \rev{work}, we will continue to use the $\{Q_n, Q_p, x_{0}, y_{0}\}$ basis with the recognition that the model can always be reformulated using another set of four basis states using Equations (\ref{eq:x100},\ref{eq:y100}).

The goal of the model is then to identify a unique combination of model parameters $\theta$ that provides the best fit against the measured voltage data. The optimization problem can be solved by implementing an error function

\begin{equation}
    \label{eq:error}
    \mathrm{E(q : \theta)} = \underbrace{
    \Up\left(y_0 - \frac{q(y)}{Q_p}\right) - \Un\left(x_0 + \frac{q(x)}{Q_n}\right)}_{
    \mathrm{V_\mathrm{full}(q)}
    } - V_\mathrm{meas}(q),
\end{equation}

where $V_\mathrm{meas}(q)$ is the experimentally-measured full cell voltage curve and the domain of $q$ is now restricted to the full-cell domain spanning $(0,Q_{\mathrm{full}})$. \rev{The model returns an optimal parameter-set which minimizes an error function of the form}

\begin{equation}
    \theta_{\mathrm{opt}} = \operatorname*{argmin}_\theta \sum_{q} \left(\lambda|\mathrm{E}(q : \theta)|^2 + (1-\lambda)\left|\frac{d\mathrm{E}(q : \theta)}{dq}\right|^2\right).
\end{equation}

\rev{The first term inside the summation is the voltage error and the second term represents the differential voltage error. $\lambda \in [0,1]$ is a weight parameter, i.e. when $\lambda=1$, the contribution from the differential voltage signal is ignored.} The optimization problem can be solved using standard non-linear least-squares solvers which are available in most scientific programming languages such as Python or Matlab. Note that, in Figure \ref{fig:dvdq}, plotted full cell voltage data corresponds to the modeled data, and the measured voltage data is omitted for simplicity. A real-world comparison of measured versus modeled full cell voltages is available in Figure S1. 

Variations the optimization method exist in the literature. For example, \cite{Lee2020a} set $\lambda=1$ and implemented additional constraints based on full cell voltage limits, i.e. $V_\mathrm{max} = \Up(y_{100}) - \Un(x_{100})$ and $V_\mathrm{min} = \Up(y_0) - \Un(x_0)$. With this approach, the model-predicted full cell capacity is guaranteed to match the measured value. However, solution convergence \rev{becomes more challenging in the presence of these additional constraints} which could make the solution space ill-conditioned. \cite{Dahn2012} included the differential voltage (dV/dQ) curves in the error function to assign preferential weight the phase transition features (Figure \ref{fig:dvdq}B) and to negate the effects of impedance mismatch between the full cell and coin cell form factors. However, \rev{these dV/dQ methods reported previously are generally not fully automated, often requiring manually adjusting model parameters through graphical user interfaces} \citep{Dahn2012} and evaluating goodness-of-fit by visually inspecting peak alignments. Overall, a ``one-size-fits-all`` optimization scheme is not known to exist yet. The specific details of the optimization may therefore need to be tuned to perform optimally for a given system.

\subsection{Model Extension: Accounting For the Inaccessible Lithium Problem}
\label{sec:inaccessible}

So far, we have laid the groundwork for the differential voltage analysis method. In this section, we outline a fundamental challenge with interpreting and reproducing model outputs known as the ``inaccessible lithium problem.'' This issue was alluded to by \cite{Truchot2014-aq} and recently expanded upon by \cite{Lu2021d}. \rev{In this section, we first describe how this issue affects interpretation of the model outputs in a manufacturing context. We then propose corrections to the differential voltage analysis model to clarify the interpretation of the model outputs.}

\rev{The ``inaccessible lithium problem'' is summarized as follows.} The \rev{model state equation} assumes that $\Up$ and $\Un$ are defined for the entire range of lithium stoichiometries spanning $y\in[0,1]$ and $x\in[0,1]$, respectively. Yet, half-cell potential curves are generally experimentally unattainable over the entire range of stoichiometries. For example, for layered oxide \rev{positive electrode systems such as NMC, full delithiation would render the host material thermodynamically unstable, impacting the reversibility of lithium intercalation reaction}. Full lithiation of the layered oxide particles is also experimentally challenging due to massive kinetic limitations towards the fully lithiated state \citep{Phattharasupakun2021-gz}. \rev{These kinetic limitations explain why it is so difficult to ``fully discharge'' a positive electrode half cell; even when discharging at small C-rates, the potential tends to rebound above 3V vs Li/Li$^+$ after resting. 

Near-equilibrium potential data on layered-oxide positive electrodes can only be feasibly obtained within a restricted stoichiometry window, which is determined by the voltage limits of coin cell data collection. Defining $\ymin$ and $\ymax$ as the minimum and maximum observable lithium stoichiometry in the positive electrode, it must be the case that $\ymin > 0$ and $\ymax < 1$.} An analogous situation applies to the graphite negative electrode, with $\xmin > 0$ and $\xmax < 1$, though the resulting errors from assuming $\xmin \approx 0$ and $\xmax \approx 1$ are generally understood to be more benign, since lithium transport in graphite is generally more facile.

To proceed with \rev{the analysis, practitioners} must either implicitly or explicitly assume that the tested range of potentials (e.g. 3.0V to 4.3V for layered oxide \rev{positive electrodes}) correspond to the full range of lithium stoichiometries, i.e.

\begin{equation}
    \label{eq:big-assumption}
    (\xmin, \xmax, \ymin, \ymax) \approx (0, 1, 0, 1),
\end{equation}

and proceed with the analysis method as described in Section \ref{sec:voltage-fitting-model}. However, this simplification will lead to errors in the model outputs. To demonstrate the magnitude of this error, we compare the \rev{positive and negative electrode capacities calculated using differential voltage analysis from Section} \ref{sec:application1} \rev{to the true electrode capacities, which can be calculated by combining the theoretical capacities with the cell design parameters, according to:}

\begin{equation}
    \label{eq:loading}
    Q = m \cdot f \cdot n_\mathrm{faces} \cdot A \cdot Q_{\mathrm{theor}},
\end{equation}

where $Q$ is the electrode capacity in Ah, $m$ is the electrode areal loading target in g/cm$^{2}$, $f$ is the mass fraction of active material in the electrode, $n_\mathrm{faces}$ is the number of active electrode faces, $A$ is the electrode area per face, excluding overhanging regions, in cm$^2$, and $Q_{\mathrm{theor}}$ is the theoretical capacity of the material in Ah/g. 

The resulting calculations, summarized in Table \ref{tbl:capacity-error}, show that the estimated \rev{positive electrode} capacity is only 54\% to 55\% of the true \rev{positive electrode} capacity, while the estimated \rev{negative electrode} capacity is 87\% to 89\% of the true \rev{negative electrode} capacity. This result was demonstrated on both the \cite{Mohtat2021b} and the \cite{Weng2021} datasets, \rev{suggesting some level of generality.} The gross underestimation of \rev{positive electrode} capacity can be primarily attributed to the fact that the layered oxide $\nmcone$ \rev{positive electrode} material used in these works retain significant lithium inventory even above the coin cell upper cut-off potential of 4.3V vs Li/Li$^+$\rev{, and consequently, $\ymin\gg 0$.}

\begin{table}[!htbp]
\centering
\begin{tabular}{r|c|c|c|c}
\toprule
{} & \multicolumn{2}{c}{\textbf{Positive Electrode}} & \multicolumn{2}{c}{\textbf{Negative Electrode}} \\
Source & Mohtat2021 & Weng2021 & Mohtat2021 & Weng2021 \\
\midrule
Number of active faces & 28 & 14 & 28 & 14 \\
Area per face (cm$^2$) & 79.20 & 79.20 & 79.56  & 79.56 \\
Loading (mg cm$^{-2}$)  & 18.50 & 17.23 & 8.55 & 7.85 \\
Theoretical capacity (mAh g$^{-1}$)  & 279.5 & 279.5 & 372 & 372 \\
Active material fraction  & 0.94 & 0.94 & 0.95 & 0.97 \\
Total capacity (Ah)  & 10.78 & 5.02 & 6.73 & 3.16 \\
\midrule
\textbf{(A)} Capacity, theoretical (mAh cm$^{-2}$) ($Q$) & 4.86 & 4.53 & 3.02 & 2.83\\
\textbf{(B)} Capacity, from fitting (mAh cm$^{-2}$) ($\tilde{Q}$) & 2.66 & 2.46 & 2.70 & 2.46 \\
\midrule
Percent of theoretical capacity observed ($\tilde{Q}/Q$) & 55\% & 54\% & 89\% & 87\% \\
\bottomrule
\end{tabular}
\caption{\textbf{Comparison of (A) electrode theoretical areal capacities versus (B) practical areal capacities derived from the differential voltage analysis.} The capacities predicted by the differential voltage analysis are lower than the theoretical values due to the `inaccessible lithium problem' \citep{Lu2021d}.}
\label{tbl:capacity-error}
\end{table}

\rev{We thus conclude that ignoring the inaccessible lithium problem} (Equation \ref{eq:big-assumption})\rev{ will likely lead to inaccurate reporting of true electrode capacities and stoichiometries, especially for layered oxide positive electrode materials such as NMC. This issue could affect the ability to directly use modeled electrode capacities to convert into electrode loadings using Equation }\ref{eq:loading}. Unfortunately, the inaccessible lithium problem is a fundamental limitation of the coin cell data collection process used to initialize $V_\mathrm{pos}$ and $V_\mathrm{neg}$ in the model.

\rev{Knowing that the inaccessible lithium problem can lead to large errors in the model outputs ($x, y, Q_n, Q_p$), we reformulate the model by defining `tilde' variables ($\tilde{x}$, $\tilde{y}$, $\tilde{Q}_n$, $\tilde{Q}_p$), to represent estimated parameters. The state equation can then be re-written as}:

\begin{equation}
    \label{eq:voltage-tilde}
    V_{\mathrm{full}}(q) = \Up(\tilde{y}) - \Un(\tilde{x})
\end{equation}

where 

\begin{align}
    \label{eq:xtilde}
    \tilde{x}(q) &= \tilde{x}_0 + \frac{q}{\tilde{Q}_n} \\
    \label{eq:ytilde}
    \tilde{y}(q) &= \tilde{y}_0 - \frac{q}{\tilde{Q}_p}.
\end{align}

This modified system is identical to the original system, but recognizes the fact that the modeled lithium stoichiometries and electrode capacities may, in general, be different from the true values which are not observable due to the limitations in coin cell data collection for $\Up$ and $\Un$. 

\rev{The estimated parameters can then be related to the true parameters by the following relations:}

\begin{align}
    \label{eq:qp-tilde}
    \tilde{Q}_p &= Q_p(\ymax-\ymin) \\
    \label{eq:qn-tilde}
    \tilde{Q}_n &= Q_n(\xmax-\xmin) \\
    \label{eq:y0y100tilde}
    \tilde{y}_0 - \tilde{y}_{100} &= \frac{y_0 - y_{100}}{\ymax - \ymin} \\
    \label{eq:x0x100tilde}
    \tilde{x}_0 - \tilde{x}_{100} &= \frac{x_0 - x_{100}}{\xmax - \xmin}.
\end{align}

\rev{This system of equations has more unknowns than equations, highlighting the difficulty in recovering the true parameters from the model alone.} Ideally, the true lithium stoichiometries may be experimentally measured, e.g. via inductively coupled plasma optical emission spectroscopy (ICP-OES) \citep{Kasnatscheew2016-lt}. \rev{True capacities may also be calculated using information about the electrode design} (Equation \ref{eq:loading}). Without more measurements, further assumptions are needed to fully resolve the estimated model parameters, e.g. $\tilde{x}_{100} \approx x_{100}$ and $\tilde{y}_{0} \approx y_0$.

\rev{Despite the data interpretation challenges introduced by the inaccessible lithium problem, we highlight several applications of the model outputs which remain unaffected by this problem. First, calculations of differences in electrode capacities are not affected.} This is because the error terms in Equations \ref{eq:qp-tilde} and \ref{eq:qn-tilde} cancel when taking ratios of capacities. This fact explains why literature reports of losses of active material (LAM) and our analysis of electrode capacity loading differences from Section \ref{sec:loadings} are still valid even though the inaccessible lithium problem may render absolute values of $Q_n$ and $Q_p$ inaccurate. \rev{Second, the inaccessible lithium problem also does not impact the process of searching for an optimal set of model parameters (Equation} {\ref{eq:error}}) since the domain for optimization is strictly only within the full cell capacity window, i.e. $q \in (0, Q_\mathrm{full})$. Missing data beyond the limits of the observable full cell voltage data is therefore inconsequential to the optimization. 

Overall, the inaccessible lithium problem effectively introduces an optional post-processing step in which the model output parameters may be converted into true parameters using Equations {\ref{eq:qp-tilde}}, {\ref{eq:qn-tilde}}, {\ref{eq:y0y100tilde}} and {\ref{eq:x0x100tilde}}. This final step would require additional input not provided for by the differential voltage model. Yet, even without knowing the true model parameters, the estimated model outputs remain useful for estimating differences in parameters such as electrode capacities. The main contribution of the model extension, including the introduction of the tilde variables, is thus to clarify that the model outputs must be interpreted as estimates of the true values due to the limitations in constructing $\Up$ and $\Un$ from data.

\subsection{Extended Model Outputs for Manufacturing Diagnostics}
\label{sec:manufacturing-dvdq}

This section defines \rev{and discusses} several additional electrochemical features relevant for battery manufacturing process control: the lithium consumed during formation ($\qsei$) the practical negative-to-positive ratio ($\nprp$) and the total cyclable lithium inventory ($\qli$). These \rev{features} can be calculated directly from the optimized parameters of the modified model $\tilde{\theta}_{\mathrm{opt}} = \{\tilde{Q}_n, \tilde{Q}_p, \tilde{x}_0, \tilde{y}_0\}$. In light of the previous discussion on the ``inaccessible lithium problem'', we make a distinction between estimated model parameters, denoted with tildes, and true model parameters. To aid understanding, expressions in this section will be given for both true values and estimated values, where appropriate.

\subsubsection{Lithium Consumed During Formation}

\rev{The capacity of }lithium consumed during formation, $\qsei$, is analogous to the first cycle Coulombic efficiency metric \citep{Mao2018-vv}, defined as ratio of discharge and charge capacity during formation. However, unlike Coulombic efficiency, which requires slow-rate charge-discharge cycles to calculate, $\qsei$ provides a consistent measure of lithium consumption even when the formation protocol consists of complex charge-discharge cycles involving multiple C-rates and partial SOC windows, as is the case with modern formation protocols \citep{An2017, Wood2019-uu}. Conveniently, since $\qsei$ can be obtained through the analysis of voltage data after formation completes, this metric enables the recovery of information about lithium consumed during formation even if formation data is missing.

We define the estimated capacity of lost lithium inventory due to formation, $\tilde{Q}_\mathrm{SEI}$, as

\begin{align}
    \label{eq:qsei}
    \tilde{Q}_\mathrm{SEI} &= q(\xmin) - q(\ymax) \\
                           &= q(\tilde{x} = 1) - q(\tilde{y} = 0)
\end{align}

\rev{This formula can be understood by considering} that, before formation, the positive electrode stoichiometry is at its highest value, $y_{\mathrm{max}}$, since all of the cyclable lithium has not yet left the positive electrode. For the same reason, the negative electrode stoichiometry is at its lowest value, $x_{\mathrm{min}}$. After formation completes, the lithium lost to the SEI will not return to the positive electrode \rev{during discharge}, causing the highest possible stoichiometry value in the positive electrode to decrease by $q(\xmin) - q(\ymax)$. Equation \ref{eq:qsei} can alternatively be interpreted as the capacity of unoccupied lithium sites in the \rev{positive electrode} when the \rev{negative electrode} is fully delithiated. 

We note that the calculation assumes that the \rev{positive electrode} voltage curve and capacity is unchanged during formation. Previous studies have shown that this assumption may not be true \citep{Kang2008-aw}, which would introduce another error contribution to this calculation which warrants further studies. From a manufacturing standpoint, if the \rev{positive electrode} system is known to be the same from batch-to-batch, then $\qsei$ may still be used to detect differences in the SEI formation process. 

\subsubsection{Re-Interpreting the Negative to Positive Ratio}
\label{sec:npr}

The negative to positive ratio (NPR) is a common term used in battery design and generally refers to the ratio of negative to positive electrode capacities. The optimization of NPR is sometimes referred to as `capacity balancing' \citep{Reuter2019-gq}, not to be confused with the process of balancing series-connected cells. Conventional wisdom suggests that the NPR be must greater than \rev{one} to prevent over-lithiation of the negative electrode during charging. Increasing NPR also implies less utilization of the \rev{negative electrode}, which could be advantageous in silicon-containing systems where volume expansion is a major contributor to cycle life degradation \citep{Luo2022-bg}. However, increasing the NPR requires putting in more \rev{negative electrode} active material into the cell without increasing usable energy content, which decreases the cell energy density. Adding more negative electrode loading could also increase the surface area for SEI reactions which could be detrimental to calendar life. \rev{The NPR is therefore} a critical cell design metric that should be optimized and tracked during manufacturing.

In this section, we first review the most commonly-held definition of the NPR and identify some conceptual gaps preventing a clear interpretation of this metric. We then propose a revised, more practical definition of the NPR which enables a more physically-grounded assessment of the lithium plating risk. 

\textbf{Issues With Conventional Definitions of NPR.} To begin, consider a common definition of the NPR based on areal loadings and theoretical electrode capacities:

\begin{equation}
    \label{eq:npr-theor}
    \mathrm{NPR}_\mathrm{theor} = \frac{m_nQ_{n,\mathrm{theor}}}{m_pQ_{p,\mathrm{theor}}}.
\end{equation}

In this equation, $m_n$ and $m_p$ are the electrode active material mass loadings in grams, and $Q_{n,\mathrm{theor}}$ and $Q_{p,\mathrm{theor}}$ are the theoretical capacities in units of mAh/g. Applying this formula to the example $\nmcone|$graphite systems from \cite{Mohtat2021b} and \cite{Weng2021} yields calculated NPR values of 0.62 for Mohtat2021 and 0.63 for Weng2021 (Table \ref{tbl:capacity-error}). \rev{This result would suggest that these cells designs have undersized negative electrode capacities and are at risk for lithium plating. However, this assessment assumes that all of the theoretical capacity of the $\nmcone$ positive electrode (279.5mAh/g) can be utilized, which is not true due to the inaccessible lithium problem (Section} \ref{sec:inaccessible}). \rev{This definition of NPR, while theoretically valid, is therefore not practically useful in a manufacturing context.}

Repeating the calculation but using the estimated values $\tilde{Q}_n$ and $\tilde{Q}_p$ to define

\begin{equation}
    \label{eq:npr-conventional}
    \mathrm{NPR}_\mathrm{conventional} = \frac{\tilde{Q}_n}{\tilde{Q}_p}
\end{equation}

leads to yet another definition of NPR found in literature. \rev{This definition partly excludes the inaccessible lithium in the positive electrode since $\tilde{Q}_p$ and $\tilde{Q}_n$ are bounded by observable potentials in the coin cell data used to construct $\Up$ and $\Un$, respectively. However, this definition remains ambiguous} because the calculated value of NPR will change based on the choice of potential ranges chosen for the half-cell reference potential curves which can vary across different datasets. NPR definitions based on ($\tilde{Q}_n, \tilde{Q}_p$) therefore do not guarantee reproducibility across different authors and datasets.

\textbf{Definition of Practical NPR}. In this section, we develop a practically relevant \rev{and consistent} definition of the NPR, which we call ``Practical NPR.'' We first return to the original motivation of comparing relative loadings of \rev{positive and negative electrode} capacities from a cell design perspective, which is to determine whether the negative electrode will become over-lithiated when the cell is fully charged to 100\% SOC. A sensible boundary condition is to set the NPR equal to unity if the negative electrode is completely lithiated when the full cell is at 100\% SOC (i.e. $x_{100} = 1$). An NPR greater than 1 would then indicate that there exists some excess \rev{negative electrode} capacity at 100\% SOC, which can serve as a margin to protect again lithium plating during charge. 

Under this construction, we realize that a practical NPR definition must then be dependent on two factors: the voltage at which 100\% SOC is defined, and how much lithium was consumed during formation. The voltage at which 100\% SOC is defined becomes relevant since higher voltages requires more utilization of both the positive and the negative electrodes, thereby decreasing the excess \rev{negative electrode} capacity. How much lithium was consumed during formation becomes relevant because more lithium consumed during formation decreases the \rev{negative electrode} lithium stoichiometry in the fully charged state, $x_{100}$, which effectively increases the \rev{negative electrode} excess capacity. In fact, since SEI formation continues over the life of the cell, $x_{100}$ will continue to decrease, which will effectively increase the \rev{negative electrode} excess capacity over life. Loss of negative active material, on the other hand, could have the opposite effect by shrinking the negative electrode curve \citep{Dubarry2012}.

Overall, a practical NPR definition cannot be a static metric, but one which changes after formation and over the life of a cell based on the competition between different degradation modes such as lithium inventory loss and active material loss.

Given the above \rev{realizations}, we now define the Practical NPR as

\begin{equation}
    \label{eq:npr-pract}
    \mathrm{NPR}_\mathrm{practical} = 1 + \frac{\tilde{Q}_{n,\mathrm{excess}}}{Q_{\mathrm{full}}},
\end{equation}

where

\begin{align}
    \tilde{Q}_{n,\mathrm{excess}} &= q(\xmax) - q(x_{100})  \\
                                  &= q(\tilde{x}=1) - q(\tilde{x}_{100})
\end{align}

represents the measured excess capacity in the negative electrode when the cell is fully charged. The Practical NPR correctly accounts for the sensitivity to the full cell upper cutoff voltage window, where increasing voltage windows leads to decreasing NPRs (i.e. less protection against lithium plating). The definition is also sensitive to changes in the electrode stoichiometries ($x_{100}$) over life, and is therefore able to account for the increase in the Practical NPR over life as lithium loss shifts $x_{100}$ to lower values over life.

Applying the Practical NPR definition to the dataset from Weng2021 and Mohtat2021 yields values of 1.14 and 1.24. This calculation shows that, after formation, the Mohtat2021 cells have more \rev{negative electrode} excess capacity than the Weng2021 cells. The difference between the calculated practical NPRs can be attributed to more lithium consumed during formation for the Mohtat2021 cells (Section \ref{sec:lithium-sei}), which increased the negative excess capacity (Section \ref{sec:plating-margin}). By contrast, none of the other definitions of NPR could distinguish this fact, since those definitions do not consider the lithium lost during formation.

A \rev{summary} comparison of different NPR calculation methodologies is provided by Table \ref{tbl:npr}.

\begin{table}[!htbp]
\centering
\begin{tabular}{r|c|c|c}
\toprule
Definition & NPR$_\mathrm{theor}$ & NPR$_\mathrm{conventional}$ & NPR$_\mathrm{prac}$ \\
\midrule
Equation & $\frac{m_nQ_{n,\mathrm{theor}}}{m_pQ_{p,\mathrm{theor}}}$ & $\frac{\tilde{Q}_n}{\tilde{Q}_p}$ & $1 + \frac{ \tilde{Q}_{n,\mathrm{excess}} }{Q_\mathrm{full}}$ \\
Ref. & Eq. \ref{eq:npr-theor} & Eq. \ref{eq:npr-conventional} &  Eq. \ref{eq:npr-pract} \\
\midrule
Excludes inaccessible lithium? & N & Y & Y \\
Considers effect of lithium lost to SEI? & N & N & Y \\
\midrule
\cite{Mohtat2021b} & 0.62 & 1.02 $\pm$ 0.009 & 1.14 $\pm$ 0.009 \\
\cite{Weng2021} & 0.63 & 1.00 $\pm$ 0.014 & 1.24 $\pm$ 0.013 \\
\bottomrule
\end{tabular}
\caption{\textbf{Comparison of various definitions of negative to positive ratio (NPR).} Numerical values show mean $\pm$ standard deviation. Values were calculated based on differential voltage analysis results performed on the datasets from \cite{Mohtat2021b} and \cite{Weng2021}.}
\label{tbl:npr}
\end{table}

\subsubsection{Total Cyclable Lithium Inventory}

The total cyclable lithium inventory in the system is imperative to track over life, since lithium inventory depletion is a primary reason for capacity loss in standard lithium-ion batteries \citep{Birkl2017a, Dubarry2022-ax}. \cite{Lee2020a} proposed that the total cyclable lithium inventory can be accounted for from the equation 

\begin{equation}
    \label{eq:qli}
    \tilde{Q}_{\mathrm{Li}} = \tilde{x}_0\tilde{Q}_n + \tilde{y}_0\tilde{Q}_p.
\end{equation}

In this equation, the first term captures the lithium trapped in the negative electrode due to the full cell minimum voltage constraint. Meanwhile, the second term includes both the cyclable lithium within the full cell operating voltage window and the inventory of lithium above the full cell maximum voltage. To gain a deeper intuition into Equation (\ref{eq:qli}), we can rewrite the same equation on the basis of the shared capacity vector $q$ as

\begin{align}
    \label{eq:qli1}
    \tilde{Q}_{\mathrm{Li}} & = \underbrace{[q(\ymin) - q(y_{100})]}_{V > V_{\mathrm{max}}} + \underbrace{[q(y_{100}) - q(y_0)]}_{\mathrm{Cyclable}} + \underbrace{[q(y_0) - q(\xmin)]}_{V < V_{\mathrm{min}}} \\
    \label{eq:qli2}
    & = q(\ymin) - q(\xmin).
\end{align}

Equation \ref{eq:qli1} shows that the total lithium inventory consists of three components: the lithium available in the positive electrode above the full cell upper cut-off voltage (first term), the lithium available within the full cell voltage window (second term), and the lithium available in the negative electrode below the full cell lower cut-off voltage (third term). Note that if $\xmin \approx 0$ is assumed, then the third term drops out and $Q_{\mathrm{Li}}$ becomes simply a statement of the total lithium inventory remaining in the positive electrode after discounting for lithium lost to the SEI. By combining Equations (\ref{eq:qsei}) and (\ref{eq:qli2}), it can be verified that 

\begin{equation}
    \label{eq:qp_from_li}
    \tilde{Q}_p = \tilde{Q}_{\mathrm{Li}} + \tilde{Q}_{\mathrm{SEI}},
\end{equation}

which clarifies the fact that all of the lithium inventory in the system originates from the positive electrode.

\subsubsection{Note on Degradation Diagnostics: From Absolute Capacities to Capacity Losses}

The differential voltage analysis method may be repeated on full cell voltage datasets collected over the course of a cycle life or calendar aging test for the aged cell system. Taking $\{\tilde{Q}_n', \tilde{Q}_p', \tilde{Q}_\mathrm{Li}'\}$ to be model outputs from the aged cell, the following familiar quantities can be defined \citep{Sulzer2021b}:

\begin{align}
    \label{eq:lamne}
    \lamne &= 1 - \frac{\tilde{Q}_n'}{\tilde{Q}_n} \\
    \label{eq:lampe}
    \lampe &= 1 - \frac{\tilde{Q}_p'}{\tilde{Q}_p} \\
    \label{eq:lli}
    \lli &= 1 - \frac{\tilde{Q}_\mathrm{Li}'}{\tilde{Q}_\mathrm{Li}},
\end{align}

where $\lampe$ is the loss of active material in the positive electrode, $\lamne$ is the loss of active material in the negative electrode, and $\lli$ is the loss of lithium inventory. Conveniently, these equations are equally valid for both true model parameters (e.g. $Q$) and observed model parameters (e.g. $\tilde{Q}$), and therefore, the `inaccessible lithium problem' introduced in Section \ref{sec:inaccessible} does not affect the numerical outcome of the study of capacity losses over the lifetime of a cell. 

Finally, we note that the definition of lithium inventory loss given by Equation \ref{eq:lli} does not provide any information on where the lithium is physically lost. It is generally understood that a primary pathway for lithium loss is through electrolyte reactions at the negative electrode to form the SEI. However, lithium can also be irreversibly trapped in the negative and positive electrodes due to electrical isolation of particles, which could occur either due to physical fracturing of active material particles \citep{Yang2012, Zhou2019, Liu2021c} or islanding of entire particles due to loss of adhesion to the binder material \citep{Muller2018-lz}. A more general expression for the loss of lithium inventory can be expressed as the sum of lithium trapped in each component:

\begin{equation}
    \mathrm{LLI} = \mathrm{LLI}_{\mathrm{pos}} + \mathrm{LLI}_{\mathrm{neg}} + \mathrm{LLI}_{\mathrm{SEI}} + \mathrm{LLI}_{\mathrm{plating}}.
\end{equation}

In this equation, the first two terms represent lithium trapped inside lithiated but electrically-isolated positive and negative electrode particles, \rev{the third term represents the lithium trapped in the negative electrode SEI} \citep{Sulzer2021b}, \rev{and the last term represents dead lithium lost to lithium plating} \citep{Yang2017-dy}. It is important to recognize that \rev{ differential voltage analysis} analysis cannot be used to decompose $\qli$ or \rev{$\mathrm{LLI}$ into their constituent parts. Differential voltage analysis thus only reveals the quantity of lithium lost without providing information about where the lithium was lost.} This limitation is fundamental to the technique, since different breakdowns of $\qli$ yield identical voltage features \citep{Birkl2017a}. More advanced, materials-level characterization, will be needed to fully understand the sources of lithium loss.

\section{Discussion}

\subsection{Data Collection Considerations}
\label{sec:datacollection}

\reva{Collecting high-quality lab data is a starting point for ensuring that the differential voltage analysis results are reproducible. We therefore dedicate this section to discussing basic experimental considerations to enable reproducible differential voltage analysis for a given cell chemistry, using Figures} \ref{fig:full-cell} and \ref{fig:half-cell} as guides.

\subsubsection{Obtaining Full-Cell Near-Equilibrium Potential Curves}
\label{sec:fullcell}

Differential voltage analysis requires the cell voltage curves to be collected at near-equilibrium conditions so that kinetic effects may be ignored. Without an assumption of near-equilibrium, Equation \ref{eq:voltage} would require additional terms to capture overpotential effects, complicating the analysis. A natural question that follows is ``at what C-rate does the voltage data needed to be collected at to constitute near-equilibrium conditions?'' This question is especially salient in battery manufacturing contexts, where the speed of diagnostic tests need to be as fast as possible. 

In work by previous authors, C-rates ranging between C/20 and C/30 are most commonly used \rev{to obtain voltage curves at near-equilibrium conditions}. For example, \cite{Dubarry2020-ug} used C/25 curves for both differential voltage analysis and incremental \rev{capacity} analysis, \cite{Lu2021d} used C/30 curves, and \cite{Keil2016a} used a fixed current value which translated into C-rates ranging between $\sim$C/10 and $\sim$C/30, depending on the cell type.

\begin{figure}[ht!]
\begin{center}
\includegraphics[width=\linewidth]{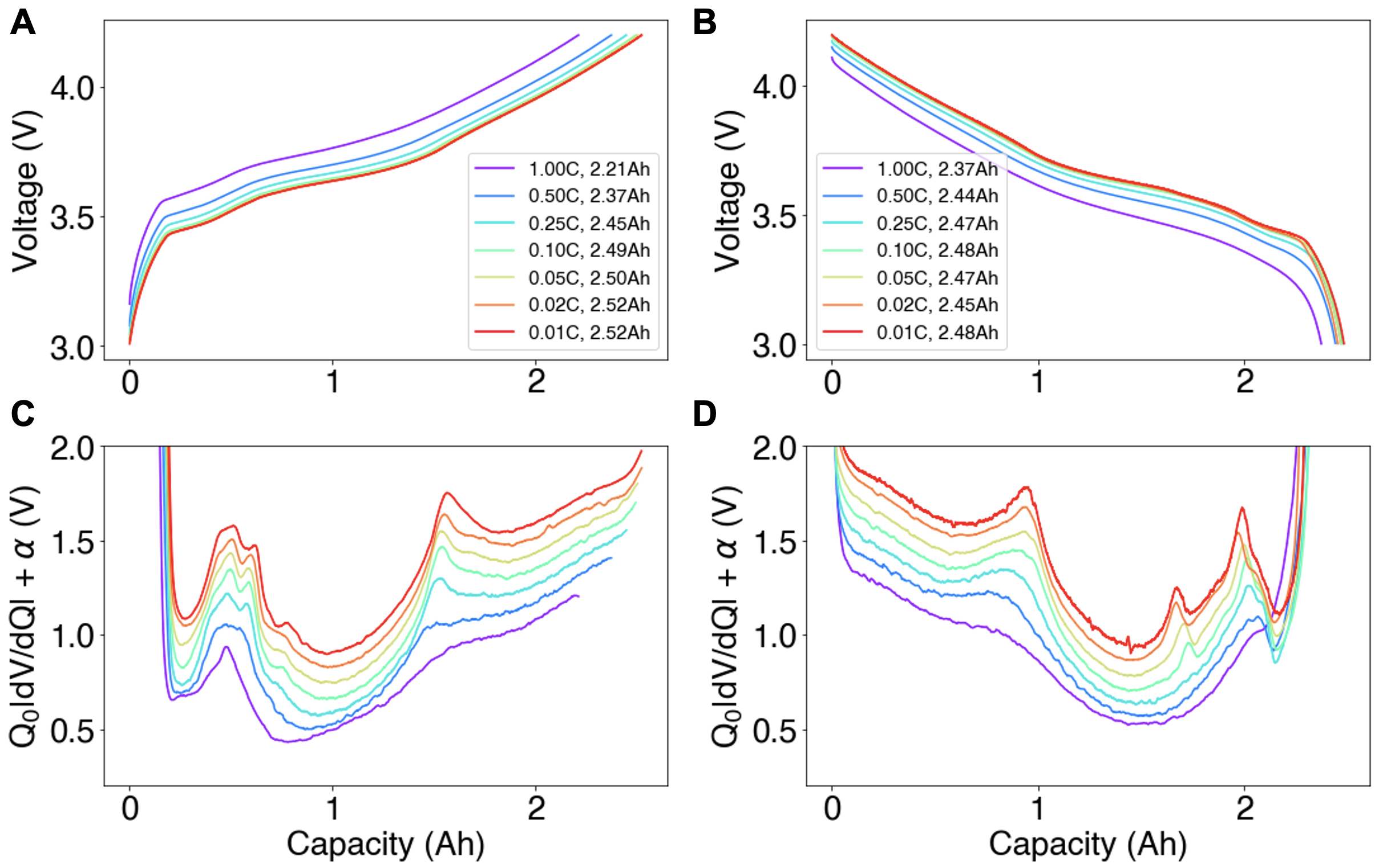}
\end{center}
\caption{\textbf{Full cell data collection example for differential voltage analysis.} (A, C) Full-cell constant-current charge curves collected at C-rates ranging between 1C and C/100 and their corresponding differential voltage curves. (B, D) The same curves, but collected on discharge. $Q_0$ is the nominal cell capacity (2.5 Ah). An arbitrary offset $\alpha$ has been added to each dV/dQ curve for clarity.}
\label{fig:full-cell}
\end{figure}

Using our own datasets, we studied the effect of C-rate on the shape of the voltage and differential voltage curves in Figure \ref{fig:full-cell}. Panel A shows an example charge dataset collected on an $\nmcsix|$graphite system, with C-rates ranging between 1C and C/100. The result shows that the C/20 charge rate achieves a capacity of 2.50Ah, which is within 1\% of the capacity of 2.52Ah measured at C/100. The corresponding differential voltage curves, shown in Panel C, shows that the C/20 rate provides peak features that are closely aligned with the near-equilibrium limit at C/100. \rev{(Note that, in this panel, each line is offset by an arbitrary constant $\alpha$ for clarity.)} Therefore, in our dataset, C/20 appeared to be a reasonable C-rate to capture a near-equilibrium condition.

While C-rates ranging between C/20 and C/30 appear to provide viable input datasets for past \rev{work}, this result may not always be true for all cell types in manufacturing. Notably, cells having higher energy densities may experience more kinetic limitations, and cells having larger form factors could experience more electrode-level lithium concentration heterogeneities. In both cases, even slower C-rates may be required to resolve the features in the differential voltage plots. Work by \cite{Mohtat2020-zp} further suggest\rev{ed} that the performance of voltage fitting methods deteriorates at higher C-rates due to peak smearing. Overall, \rev{more work is needed to quantify the effect of C-rate on the analysis outputs as well as the generalizability of C-rate setpoints across different cell chemistries and cell designs.}

\rev{The directionality of charge must also be considered.} Figure \ref{fig:full-cell}B,D show that the voltage and dV/dQ features on discharge are, similar, but not identical, to those on charge. At C/100, the low-SOC peaks, which correspond to graphite features, appear to have sharper peaks on discharge. An additional peak also appears in the discharge data, which can be attributed to the graphite 2L staging reaction (see Figure \ref{fig:half-cell}C). \rev{Practically speaking, the presence of the charge-discharge asymmetry suggests that the differential analysis outputs may be different depending on whether the charge or discharge curves are used. The extent of these differences should be further studied, which we leave as future work.}

\subsubsection{Obtaining Half-Cell Near-Equilibrium Potential Curves}
\label{sec:halfcell}

Electrode near-equilibrium potential curves can either be obtained experimentally \rev{from scratch using coin cell testing} \citep{Hu2021-pf, Murray2019-vf} or downloaded from open-access databases such as LiionDB \citep{Wang2022-zo}. In either case, due caution is necessary for several reasons. 

First, the majority of datasets report half-cell potentials versus lithium stoichiometries, but since there is no common voltage range setting, different half-cell datasets will lead to different modeled outputs for electrode capacities and stoichiometries (see Section \ref{sec:inaccessible}). 

Second, the experimental procedures used to obtain the curves may differ, with some authors using continuous currents, while other authors obtaining `quasi-static', or `pseudo-OCV' curves, by combining data collected after some voltage period at each SOC, similar to the galvanostatic intermittent titration (GITT) technique \citep{Chen2020-tg}. While the latter method could theoretically provide a more accurate near-equilibrium potential curve, fewer data points are typically collected to ensure the test finishes within a reasonable time. As a result, some characteristic features related to phase transitions in the negative electrode (Figure \ref{fig:full-cell}C) may be lost \citep{Hess2013-zt}.

\begin{figure}[ht!]
\begin{center}
\includegraphics[width=\linewidth]{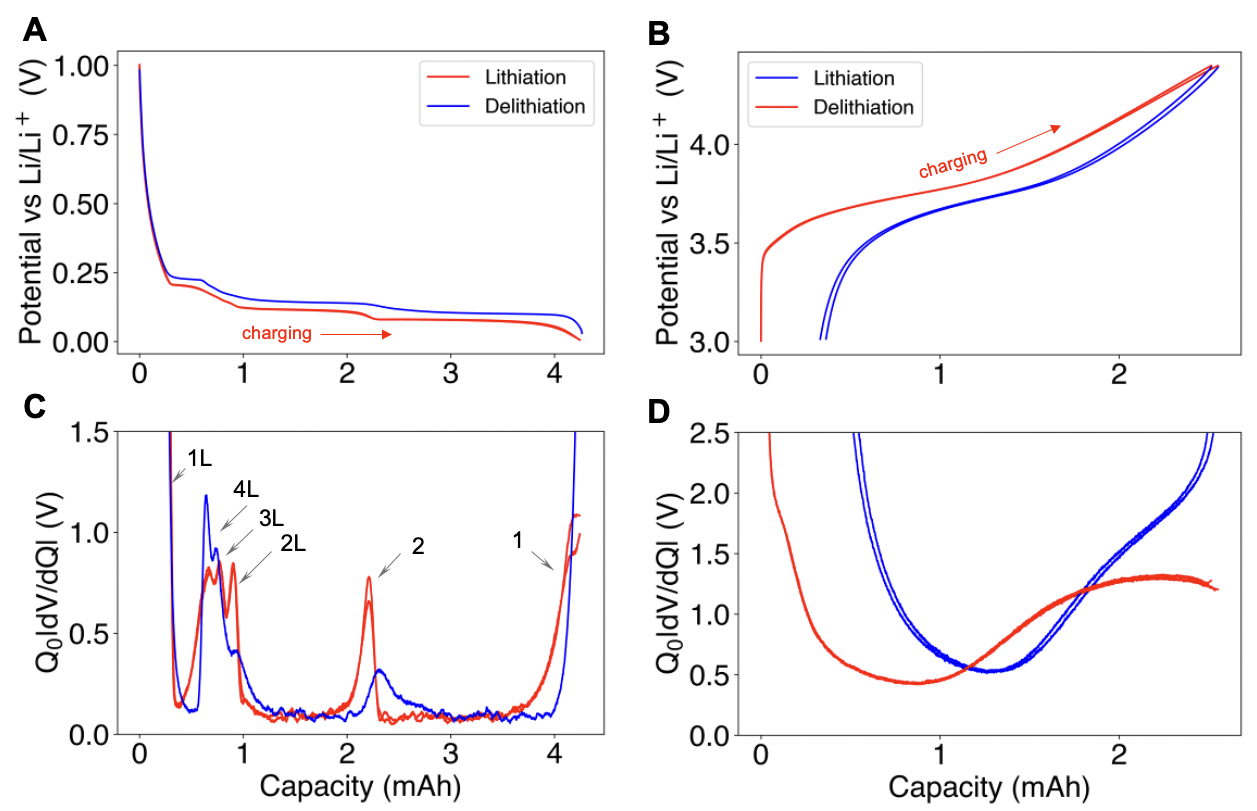}
\end{center}
\caption{\textbf{Half-cell data collection example for differential voltage analysis.} (A,C) Graphite$|$lithium half-cell lithiation and delithiation curves and their corresponding different voltage curves. (B,D) $\nmcone|$lithium half-cell lithiation and delithiation curves and their corresponding differential voltage curves. All data is measured at C/50. To indicate reproducibility, data from two consecutive charge-discharge curves have been provided (except for the graphite delithiation curve). The direction corresponding to full cell charging is indicated by a red arrow.}
\label{fig:half-cell}
\end{figure}

A third reason for exercising caution when using half-cell data obtained from literature is that the current direction is sometimes not reported, yet current direction materially impacts the features seen in the voltage curves. Figure \ref{fig:half-cell} shows that the current direction influences the voltage (and differential voltage) features for both a graphite negative electrode (Panels A,C) and an $\nmcone$ \rev{positive electrode} (Panels B,D). At the negative electrode, a hysteresis gap between the lithiation and the delithiation curves exists, which persists across consecutive charge-discharge cycles. The characteristic differential voltage feature occurring at $\sim$2.3 Ah, corresponding to the stage 2 transition, is sharper and higher in magnitude in the lithiation direction. The staging reactions 2 through 4 are also distinctly different. In particular, the staging reaction 2L appears sharper on lithiation and the stage 4L reaction appears sharper on delithiation \citep{Hess2013-zt}.

At the positive electrode, a large kinetic limitation is observed at the end of the lithiation curve, causing the lithiation capacity to be measurably lower than the delithiation curve across multiple cycles. The observation of poor kinetics during lithiation of the positive electrode has been experimentally confirmed by \cite{Kasnatscheew2016-lt} and thoroughly discussed in \cite{Phattharasupakun2021-gz}. 
Overall, charge-discharge asymmetry exists at the material level for both the positive and the negative electrodes. For high-fidelity \rev{differential voltage} analysis, the current direction used to obtain the electrode potential curves \rev{should be aligned} against what is used for the full cell. Doing so will improve the model's ability to fully describe the full cell \rev{voltage} curve using the \rev{electrode} potential curves.

\subsubsection{Data Logging Frequency and Data Smoothing}

The data logging rate and filtering method will affect the smoothness of the collected voltage curves and its derivatives. It is generally recommended to over-sample than under-sample to minimize data loss. Noise in the over-sampled data can be overcome by post-processing the data using filtering methods. A common approach is to use a Savitzky-Golay filter \citep{Savitzky1964-td} which is included as part of most scientific programming languages. Note, however, that the process of selecting filtering parameters, such as window size and polynomial order, may introduce distortions in the data, decreasing the reproducibility of modeled results \citep{Feng2020-yq, Lu2021d, Schmid2022-ba}. To improve model reproducibility, more studies are needed to understand the degree to which data filtering strategies impact model outputs, which we leave as future work.

\subsection{Comments on Factory Deployment}
\label{sec:deployment}

Here, we \rev{summarize} how a battery manufacturer might deploy the differential voltage analysis method in the battery factory for online process control and quality control applications. \rev{We also highlight the remaining knowledge gaps that may prevent deployment today which warrant further research.}

\rev{Enabling the differential voltage analysis in manufacturing requires both} half-cell and full-cell data collection. The half-cell data only needs to be collected once. The data, once vetted, can be stored as part of the model input parameters and be re-used to analyze all full cell data belonging to the same family of \rev{positive and negative electrode} material sets, e.g. across multiple cell batches and production lines.

\reva{The largest barrier to method deployment is then in the acquisition of the full cell near-equilibrium data. As discussed previously, current methods for differential voltage analysis generally require around 20 hours in order to collect the full cell voltage trace. If the measurement method is to be deployed on every cell coming out of formation, then the production throughput will decrease.} Considering that the industry trend is to decrease formation times \citep{Wood2019-uu}, any additional process that will make the total time on test longer may be untenable. 

\reva{To improve the adoption of the method, we propose several pathways. First, rather than collecting the full cell voltage data after formation, the voltage data during the formation process itself may be processed for differential voltage analysis. To our knowledge, this method has yet been reported in academic literature, and is the subject of future work.}

Second, a study should be conducted to understand the influence of higher C-rates on the modeled parameters. Historically, some of the challenges with differential voltage analysis arose in the analysis of aged cells, where the effective current density increases as the cell capacity decreases. For example, at 50\% capacity retention, a ``C/20'' charge step would only take approximately 10 hours to complete since the C-rate is defined with respect to the nominal (pristine) cell capacity. In battery manufacturing, we are only concerned with studying the electrochemical state of the pristine cells without consideration for aging. Hence, the C-rates traditionally thought to be necessary to enable \rev{differential voltage} analysis may be too conservative in a manufacturing context. A future study of the influence of C-rate on the modeled electrochemical parameters will provide some necessary guidance in this direction. 

Finally, not every cell may need to be fingerprinted. For example, retaining and measuring several dozen cells (out of hundreds of thousands built per day) could provide enough statistical information to track day-to-day changes in electrochemical process parameters. These measurements can still be made on formation equipment, and would only require a small, dedicated number of formation cyclers, to be withheld for this purpose. This partitioning method can also encourage the adoption of more advanced characterization methods such as \rev{computed tomography} (CT) \citep{Wu2018-bx, Bond2022-xh, Gauthier2022-xt}, ultrasound imaging \citep{Bommier2020-so, Deng2020-kd}, and strain measurements \citep{Mohtat2022-xi}. Measuring all of these parameters on the same cell could enable a more complete perspective on the cell electrochemical and mechanical state, enabling deeper insights. 

\subsection{Improving Method Reproducibility}
\label{sec:reproducibility}

Besides logistical challenges of deployment, another major and perhaps larger obstacle preventing adoption of the differential voltage method is the difficulties with implementation and interpretation of model outputs. These difficulties stem from a lack of literature guidance on a standard, rigorous method to implement the fitting model and an absence of understanding of how sensitive the model outputs are to small differences in input datasets, including both full cell and half cell data. This work sought to clarify the model interpretation by enabling a quantified understanding of model errors due to the inaccessible lithium problem (Section \ref{sec:inaccessible}). We also discussed elementary data collection considerations needed to enable a working model (Section \ref{sec:datacollection}).

Several avenues may be taken to further improve reproducibility of the electrochemical features derived from the \rev{differential voltage analysis method.} First, model input sensitivities need to be better understood. \rev{This work introduced several input sensitivity considerations} including C-rate and current direction (charge versus discharge). However, a discussion on sensitivity to temperature \rev{was not covered by this work. Moreover, a quantitative assessment of the input-output sensitivities have not yet been completed.} A logical next step is \rev{thus} to quantify model output differences due to these differences in model inputs. 

\reva{Another important consideration involves understanding the optimizer's ability to find the optimal solution $\theta_\mathrm{opt}$. Reproducibility requires that there exists a global minimum value of the error function defined by Equation }\ref{eq:error} \reva{and that the optimizer can converge to this value. However, in general, neither condition is guaranteed since the objective function is, in general, non-convex. The solution uniqueness for the positive electrode parameters $\tilde{Q}_p$ and $y_0$ may be especially challenging for NMC systems which are relatively featureless} (Figure \ref{fig:dvdq}). Some past efforts have been made to understand how model output parameter uncertainty increased as the data collection voltage window decreased \citep{Lee2020-jy}. These studies should be expanded. \rev{A better understanding here will enable practitioners to quantify} how much of the variability in model-predicted outputs, such as those shown in Figure \ref{fig:correlations-manufacturing}, are due to true manufacturing variability versus model uncertainty.

Finally, it must be recognized that all of the electrochemical features extracted using the differential voltage model are estimates of the true values (Section \ref{sec:inaccessible}). Experimental verification of model-predicted outputs, such as \rev{positive electrode} capacity, \rev{negative electrode} capacity, lithium lost to SEI, cyclable lithium inventory, and negative to positive ratios, can be challenging to obtain, but may be necessary to obtain confidence in the model-predicted outcomes.


\section{Conclusion}

This work clarified how \rev{the differential voltage analysis method} can be automated to improve online battery manufacturing process control. Using an example manufacturing dataset, we demonstrated how \rev{modeled} outputs, such as \rev{positive and negative electrode capacities, lithium lost to form the SEI, and negative to positive ratios, can be used to gain insight into the sources and consequences of} manufacturing variability. In our example, we \rev{found} that the source of variability in full cell capacity for one batch of cells is due to variations in \rev{positive electrode} loadings. We also detected a three-fold difference in the amount of lithium consumed consumed to the SEI between two different cell batches, a conclusion \rev{we were able to make} without observing any data from the formation cycles or using any destructive analytical methods.

To facilitate adoption of the \rev{differential voltage analysis method} in manufacturing, we detailed the mathematical formalism required to implement the method as well as the input data requirements. We extended the base model formulation to explicitly account for model parameter errors due to the inaccessible lithium problem. Using the reformulated model, we defined an expanded set of electrochemical features which included the lithium consumed during formation, the practical NP ratio, the excess \rev{negative electrode} capacity, and the total cyclable lithium. When combined, these features provide a ``fingerprint'' of the electrochemical state of the pristine cell which can then be used to \rev{adjust parameters in various manufacturing steps including electrode manufacturing and the formation process.}

Besides clarifying the model implementation details, \rev{this work also} recognized the importance of obtaining high-quality experimental data, which includes both full cell and half-cell data, to enable reproducible results. We identified several basic data collection considerations, including the C-rate, current direction (charge versus discharge), and data smoothing method, as being potential factors affecting model output reproducibility.

With an improved awareness of the nuances involved with model construction, model output interpretation, and data collection, the battery community can be better positioned to deploy the \rev{differential voltage analysis method in factories, driving continuous improvement in the battery manufacturing process.}


\section*{Data Availability}

This study presented datasets from multiple works. The cell test dataset from \cite{Mohtat2021b} is available at \url{https://doi.org/10.7302/7tw1-kc35}. The cell test dataset from \cite{Weng2021} are available for download at \url{https://doi.org/10.7302/pa3f-4w30}. All source code used to analyze the datasets and generate the plots are available at \url{https://github.com/wengandrew/dvdq}. The differential voltage analysis code was written in MATLAB. The data post-processing tools and data visualizations were developed in Python.

\bibliographystyle{unsrtnat}
\bibliography{main}  






\end{document}